\documentclass[11pt,a4paper]{article}
\usepackage{jheppub}
\usepackage{amsmath}
\usepackage{amsfonts}
\usepackage{amssymb}
\usepackage{braket}
\DeclareFontFamily{U}{bigshuffle}{}
\DeclareFontShape{U}{bigshuffle}{m}{n}{
  <5-8> s*[1.7] shuffle7
  <8->  s*[1.7] shuffle10
}{}
\DeclareSymbolFont{BigShuffle}{U}{bigshuffle}{m}{n}
\DeclareMathSymbol\bigshuffle{\mathop}{BigShuffle}{"001}
\DeclareMathSymbol\bigcshuffle{\mathop}{BigShuffle}{"002}

\usepackage{shuffle}
\usepackage{xcolor}
\usepackage{tikz-feynman}
\definecolor{cof}{RGB}{219,144,71}
\definecolor{pur}{RGB}{186,146,162}
\definecolor{greeo}{RGB}{91,173,69}
\definecolor{greet}{RGB}{52,111,72}
\newcommand{\Z}[0]{\mathbb{Z}}

\newcommand{\li}[0]{\text{Li}}

\title{Symbol Recursion for the dS Wave Function}
\author[1]{Aaron Hillman}
\affiliation[1]{Department of Physics, Jadwin Hall, Princeton University, NJ, USA 08540}
\emailAdd{aaronjh@princeton.edu}
\abstract{We present a recursive rule for the symbol of perturbative contributions to the vacuum wave function of a conformally coupled scalar in FRW cosmologies. The rule applies exactly for a class of interactions and cosmologies, which contains $\lambda \phi^3$ in $dS_4$, a case of particular relevance as a source of building blocks for inflationary correlators.  We use the rule to efficiently reproduce the tree-level four-point contribution and present novel computations of the bubble integrand and the tree-level five-point contribution.  Our results apply equally well to the computation of Witten diagrams in Euclidean AdS.}
\begin{document}
\maketitle\text{}\\
\tableofcontents
\section{Introduction}
The flat-space S-matrix is tightly constrained by locality and unitarity.  We understand how these constraints are manifest at tree-level and one-loop-level
, though the constraints on the space of functions that might appear at two loops or more are still not known.  Additionally, recent years have exposed the rich mathematical structures behind scattering amplitudes, such as the Grassmanian, Amplituhedron, Associahedron, and cluster algebras to name a few.  These mathematical frameworks have furnished representations of scattering amplitudes as answers to elegant questions purely in kinematic space.  In this context, locality and unitarity are derivative notions, emerging from the abstract conditions imposed in this space of kinematic data at infinity, such as positivity.  The situation in cosmology is vastly underdeveloped by comparison.  We have no significant constraints on the vacuum wave function beyond normalizability, and the stock of theoretical data is relatively meager. Moreover, the need for such a boundary description is more urgent, as our only windows on the early universe are spatial correlations in the late universe.  We have no choice but to determine how causal evolution in the bulk  has been encoded in the future boundary, how time emerges in the kinematic space at infinity.  \\
\indent The study of cosmological polytopes, positive geometries in the kinematic space of spatial momenta at infinity, was initiated in \cite{cosmopoly}, taking a necessary step in this direction.  There, perturbative contributions to the vacuum wave function are the canonical forms of certain polytopes in the space of external and edge energies of Feynman graphs.  In this setting, causal evolution and unitarity emerge as factorization properties of the underlying geometry \cite{emergent}.   A systematic cosmological bootstrap program exploiting the conformal symmetry of cosmological correlators in de Sitter was also initiated recently \cite{cosmobootstrap}.  In this context, consistent time evolution in the bulk is encoded in differential equations obeyed by correlators in kinematic space.  In both cases, the signatures of time-dependent physics emerge in a space with no explicit reference to time.  The cosmological bootstrap also demonstrated the particular importance of correlators of conformally-coupled scalars as they can be used as seeds from which the correlators for externally massless fields and other mass and spin exchanges can be deduced by acting with certain differential operators.  The wave function contributions for light external states can be found from the wave function considered here using the differential operators derived in \cite{lightstates}.  This further motivates a careful study of the conformally coupled case for reasons beyond its relative simplicity.      

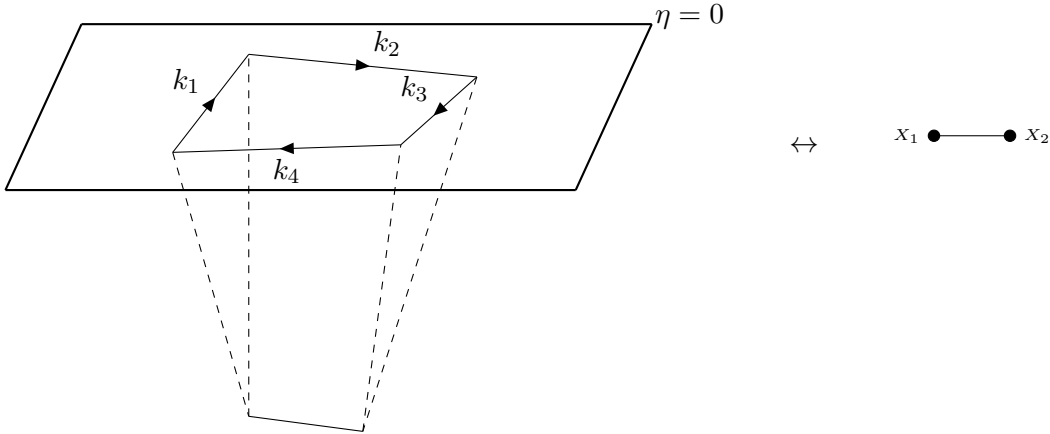
\begin{figure}[h!]
\centering
	 \begin{tikzpicture}[baseline=0mm]
  \begin{feynman}[inline=(p4)]
 	\vertex (v1) at (0, 0);
 	\vertex (v2) at (1, 1.3);
 	\vertex (v3) at (4, 1);
 	\vertex (v4) at (3, 0.1);
 	\vertex (p1) at (-2.2, -0.5);
 	\vertex (p2) at (-1.2, 1.7);
 	\vertex (p3) at (6.3, 1.7);
 	\vertex (p4) at (5.3, -0.5);
 	\vertex (i1) at (1, -3.5);
 	\vertex (i2) at (2.5, -3.7);
 	\vertex (eta) at (6.8, 1.8) {$\eta = 0$};
 	\diagram*{
 	(v1) -- [fermion, edge label=\( k_1\)] (v2) -- [fermion, edge label = \( k_2\)] (v3) -- [fermion, edge label'=\(k_3\)] (v4) -- [fermion, edge label=\( k_4\)] (v1);
 	(p1) -- [thick] (p2) -- [thick] (p3) -- [thick] (p4) -- [thick] (p1);
 	(v1) -- [dashed] (i1) -- [dashed] (v2);
 	(v3) -- [dashed] (i2) -- [dashed] (v4);
 	(i1) --  (i2);
 	};
  \end{feynman}
 \end{tikzpicture} \hspace{6mm}$\leftrightarrow$ \hspace{6mm} \begin{tikzpicture}[scale=1.5]
     \begin{feynman}
     	\vertex (v1);
     	\vertex[right=1cm] (v2);
     	\vertex[left=.05cm] (v3) {\tiny $X_1$} ;
 		\vertex[right=1.05cm] (v4) {\tiny $X_2$};
     	\diagram*{
     	(v1) -- (v2)
     	};
     	\draw[fill=black] (v1) circle(.5mm);
\draw[fill=black] (v2) circle(.5mm);
     \end{feynman}
 \end{tikzpicture}
 \caption{The dumbbell is a Feynman diagram where we have truncated the particles propagating to the boundary (dashed lines).  The vertex energies are $X_1 = k_1+k_2$ and $X_2 = k_3+k_4$ and the solid edge is the propagation of an intermediate fluctuation.}
 \label{dumbell}
\end{figure}

Part of the story of the flurry of progress in understanding scattering amplitudes was the discovery of efficient recursion relations, such as BCFW recursion, which facilitate the production of a wealth of theoretical data to be studied.  We take a modest step toward recursion relations for cosmological correlators, presenting a recursive rule for the symbol of perturbative contributions to the vacuum wave function of conformally coupled scalars. A general but somewhat heavy-handed formula for the symbol of perturbative contributions to the wave function was presented in \cite{cosmopoly} employing the methods of \cite{oneloop} which hold for arbitrary Aomoto polylogarithms built out of pairs of polytopes.  This picture, despite its generality and beautiful interpretation of the symbol as computed via ``walks" through the cosmological polytope, obfuscates the physics that is made manifest by the recursive rule.  In particular, as is familiar from the S-matrix, the locations of singularities and factorization properties of the answer on singularities is how causality and unitarity are encoded in the answer.  The situation is no different in cosmology, though we already encounter branch cuts at tree-level. \\
\indent This paper is organized as follows.  In Section \ref{background} we review the model under consideration and provide a brief review of symbol calculus.  In Section \ref{rule} we state the recursive rule for the symbol of perturbative wave function contributions and provide a derivation directly in terms of time integrals as well as illustrating the interpretation in terms of the cosmological polytope.  We then illustrate the conceptual clarity and efficiency of the rule in Section \ref{examples} by computing the symbols for the two-site chain (Fig. \ref{dumbell}), two-site loop, and the three-site chain.  In Section \ref{integration} we fully integrate these examples reproducing the result of \cite{cosmocollider} at four-points and providing novel results for the five-point tree-level contribution and the bubble integrand. 
\section{Background}\label{background}
\subsection{The Model}
We consider a conformally coupled scalar in a general FRW background as studied in \cite{cosmopoly, emergent, smatrixtowavefuntion}.  We work in $d+1$ dimensions where $d$ is the number of spatial dimensions.  The action is  
\begin{align}
	S = \int d\eta d^dx \sqrt{-g}\left( \frac{1}{2} g^{\mu\nu}\partial_\mu \phi \partial _\nu \phi + \xi R\phi^2+\sum\limits_k \frac{\lambda_k}{k!}\phi^k \right)
\end{align}
With
\begin{align}
	\xi = \frac{d-1}{4d} \hspace{20mm} ds^2 = a(\eta)^2 (d\eta^2-dx^2)
\end{align}
where we use conformal time $\eta \in (-\infty, 0]$ and the choice of $\xi$ sets the scalar to be conformally coupled.  That is,
this action is conformally equivalent to the following flat-space action 
\begin{align}
	S[\phi] = \int d\eta d^dx\left( \frac{1}{2}(\partial \phi)^2+\sum\limits_k \frac{\lambda_k(\eta)}{k!}\phi^k\right)
\end{align}
Under the transformation 
\begin{align*}
	g &\to a(\eta)^{-2}g\\\\
	\phi &\to a(\eta)^{-\Delta}\phi
\end{align*}
where $\Delta = \frac{d-1}{2}$ for a scalar.
We now have time-dependent couplings given by
\begin{align}
	\lambda_k(\eta) = \lambda_k[a(\eta)]^{2+\Delta(2-k)}
\end{align}
We are interested in computing the vacuum wave function computed by the path integral with non-vanishing boundary conditions in the future.  
\begin{align}
\label{psi}
	\Psi_{BD}[\Phi] = \int\limits^{\phi(0) = \Phi}_{\phi(-\infty(1-i\epsilon)) = 0} \mathcal{D}\phi e^{iS[\phi]} \equiv \exp \left[i\sum\limits_n \frac{1}{n!} \int \prod\limits_i d^dk_i \Phi(k_i)\psi_n(\{ k_i \}) \right]
\end{align}
Where the $i\epsilon$ prescription for $\eta$ projects out the Bunch-Davies vacuum and the $\psi_n$ are the wave function contributions that we are interested in computing in \textit{spatial} momentum space.  These are $n$-point functions computed by the ordinary Feynman rules of the theory.  We emphasize that we mean spatial momentum space, as the times of interaction vertices are explicitly integrated over.  Each $\psi_n$ contains an overall momentum conserving $\delta$-function from spatial translation invariance and we will be computing these ``stripped" $\psi_n$'s.  We now justify the equivalence of the latter two expressions in Eq. \ref{psi}.    When computing the wave function it is convenient to split the field into a classical momentum-space mode function and a fluctuation  
\begin{align}
	\phi(\vec p, \eta) = \Phi(\vec p)e^{iE_p\eta}+\delta \phi(\vec p, \eta)
\end{align}
Note that $\Phi(\vec p)$ is the same as that appearing in the boundary conditions on the path integral over $\phi$.
 If we plug the expression above into the action, we now have a path integral over the fluctuation with the boundary conditions being that the fluctuation vanish as $\eta \to 0$ and as $\eta \to -\infty$.
By plugging the split expression in for $\phi$, expanding, and taking functional derivatives of this path integral with respect to the boundary value, which now operates like a classical source, one can verify the exponential expression.  The perturbative contributions $\psi_n$ are therefore computed using Feynman rules for the fluctuations.  In order to compute these contributions, we need  the bulk-to-bulk propagator for fluctuations and the bulk-to-boundary propagator given by the classical solution 
\begin{align}
	G_{\text{BB}}(E_e, \eta_v, \eta_{v'}) &= \frac{1}{2E_e}\left(e^{-iE_e(\eta_v-\eta_{v'})}\theta(\eta_{v}-\eta_{v'}) +e^{-iE_e(\eta_{v'}-\eta_{v})}\theta(\eta_{v'}-\eta_{v})-e^{iE_e(\eta_v+\eta_{v'})} \right)\\
	G_{\text{B}\partial\text{B}}(E, \eta) &= e^{iE\eta}
\end{align}
From now on we suppress the subscripts and $G$ refers to the bulk-to-bulk propagator.
We see that the propagator for fluctuations is the usual spatial momentum space Feynman propagator with an additional term ensuring that the fluctuations vanish at $\eta = 0$, consistent with the boundary condition in the path integral.
The subscripts on the times denote the vertices the propagator connects.  The process of computing the wave function is now reduced to an ordinary connected Feynman graph expansion.  The result of this work applies exactly to the wave function computed for interactions $\lambda_k \phi^k$ obeying
\begin{align}\label{oneovereta}
	[a(\eta)]^{2+\frac{d-1}{2}(2-k)} = \frac{1}{\eta}
\end{align}
So that $\lambda_k(\eta) = \frac{\lambda_k}{\eta}$. We can pass to Fourier space for the coupling most easily in this case, as 
\begin{align}
	\frac{1}{\eta} = -i\int\limits_0^{\infty(1-i\epsilon)} e^{i\varepsilon \eta}d\varepsilon
\end{align}
With the time-dependence of the coupling taken care of by the energy integral, we can define the energy integrand 
\begin{align*}
	\psi_\mathcal{G} = \int \prod\limits_{v \in \mathcal{V}}d\eta_v e^{iX_v \eta_v}\prod\limits_{e \in \mathcal{E}}G(Y_e, \eta_{v_e}, \eta_{v_e'})
\end{align*}
Where the $x_v = \varepsilon_v + X_v$ denote the sums of energies running to the boundary from a vertex $v$ and the $Y_e$ are the energies of edges.  These are the rational functions that can be computed using the tools in \cite{cosmopoly}.
The full wave function contributions take the form 
\begin{align}
	\psi_n(\{ X_v\}, \{ Y_e \}) = \int\limits_\mathcal{R} \prod\limits_{v \in \mathcal{V}} dx_v \psi_\mathcal{G}(\{ x_v\}, \{ Y_e \})
\end{align} 
where $\mathcal{R}$ is the rectangular region defined by $x_v > X_v$ for each vertex.  We have also suppressed powers of the coupling.    The sets in the argument of $\psi_\mathcal{G}$ are the shifted, unintegrated vertex energies in Fourier space and the $Y$'s are the fixed edge energies.  For couplings and cosmologies consistent with Eq. \ref{oneovereta} we know that, defining
\begin{align}
f_{\mathcal{G}} &= \left(\prod\limits_e 2Y_e\right)\psi_\mathcal{G}\\
F_{\mathcal{G}}(\{ X_v\}, \{ Y_e \})
 &= (-1)^{n_v}\left(\prod\limits_e 2Y_e\right)	\psi_n(\{ X_v\}, \{ Y_e \})
 \end{align}
 the function $F_{\mathcal{G}}$ will be a pure polylogarithm as $f_{\mathcal{G}}$ has unit partial residues.  The overall sign is determined by $n_v$, the number of vertices.  It is convenient to define $F_{\mathcal{G}}$ this way so that we do not worry about how the overall sign of the symbol changes depending on whether the number of vertices is even or odd.  The change in sign comes from the fact that external kinematic data are the lower limits of integration.  A comment on the diagrammatics used here and and in \cite{cosmopoly, emergent,lightstates} is in order.  The convention employed for the diagrams is that the edges running to the boundary are truncated, as these edges do not correspond to the propagation of fluctuations and so any number of propagators running to the boundary from a vertex merely contribute via the external energy $X_v$.  The process of trunctation is depicted for the four-point case in Fig. \ref{dumbell}.  \\
 \indent One concrete model in this class of theories, $\lambda \phi^3$ theory in $dS_4$, is of particular interest as it is a starting point relevant for understanding the structure of inflationary correlators \cite{cosmocollider, cosmobootstrap}.  We briefly discuss the possibility of application to more general backgrounds and interactions in Section \ref{moregeneral}.  It is also worth noting at the onset that our results are formulated in the Lorentzian language of $\Psi_{BD}$ but are just as good for Witten diagrams contributing to $Z_{EAdS}$, the partition function in Euclidean AdS. In particular, the recursive rule for contributions  to the wave function as presented in this paper can be ported to Euclidean AdS.  This relation is discussed in greater detail in \cite{nongauss}.   
\subsection{Polylogarithms and Symbols}
Here we review the necessary aspects of symbol calculus and multiple polylogarithms necessary for this work.  A complete review of these topics and more can be found in \cite{duhr}.  The symbol map for polylogarithms has multiple definitions.  Here we employ a definition considered in \cite{goncharov, duhr}.  If we have a complex-valued transcendental function  $F^{(n)}$ of pure weight $n$  which has a differential obeying the following 
\begin{align}
	d F^{(n)} = \sum\limits_i F^{(n-1)}_i d\log R_i
\end{align}
where the $F^{(n-1)}_i$ are transcendental functions of weight one less and the $R_i$ are rational functions, the symbol of the function is then defined recursively via 
\begin{align}
	\mathcal{S}(F^{(n)}) = \sum\limits_i \mathcal{S}(F^{(n-1)}_i) \otimes R_i
\end{align}
we see that the symbol inherits its linearity properties from the logarithm i.e.
\begin{align}
	a \otimes b-a\otimes c = a\otimes \frac{b}{c}
\end{align}
Moreover, the symbol of a function that is the product of two transcendental functions $F^{(n)}$ and $G^{(m)}$ of weights $n$ and $m$ respectively, has weight $n+m$ and its symbol is the ``shuffle product" of the symbols of $F$ and $G$, given by the sum of all ways of interleaving the symbol words of each, where this means interlacing the symbol entries between a pair of words while maintaining the internal order of each word.  That is 
\begin{align*}
	a_{1}\otimes \dots \otimes a_{n}\shuffle a_{n+1} \otimes \dots a_m = \sum_{\sigma \in \Sigma(n, m)} a_{\sigma(1)}\otimes \dots \otimes a_{\sigma(m)} 
\end{align*}
where we introduced the notation $\shuffle$ for the shuffle product and $\Sigma(n, m)$ is the set of all shuffles on a symbol word of length $n$ and a word of length $m$, i.e. the subset of the symmetric group $S_{n+m}$ defined by  
\begin{align*}
	\Sigma(n, m) = \left \{ \sigma \in S_{n+m} \big | \sigma^{-1}(1) < \dots < \sigma^{-1}(n)\hspace{3mm} \&  \hspace{3mm}\sigma^{-1}(n+1) < \dots < \sigma^{-1}(m)\right \}
\end{align*}
As an example 
\begin{align*}
	a\otimes b \otimes c \shuffle e\otimes f = \hspace{1mm}&a\otimes b\otimes c\otimes e\otimes f+a\otimes b\otimes e\otimes c\otimes f+a\otimes e\otimes b\otimes c\otimes f\\
	&e\otimes a\otimes b\otimes c\otimes f + e\otimes a\otimes b\otimes f\otimes c+e\otimes a\otimes f\otimes b\otimes c\\
	&e\otimes f\otimes a\otimes b\otimes c
\end{align*}
This is trivially extended to sums of words by linearity.\\
\indent The classical polylogarithms have the following recursive definition as iterated integrals 
\begin{align}
	\text{Li}_1(z) &= -\log(1-z)\\
	\text{Li}_n(z) &= \int\limits_0^z \text{Li}_{n-1}(t)d\log(t)
\end{align}
Therefore we see that the symbol of a classical polylogarithm is 
\begin{align}
	\mathcal{S}(\text{Li}_n(z)) = -(1-z)\otimes z\dots \otimes z
\end{align}
where there symbol above is an $n$-fold tensor product.
The classical polylogarithms each have a branch cut on the real axis from 1 to $\infty$.  The discontinuity across this cut is readily obtained from the symbol.  We see in the case of classical polylogarithms that 
\begin{align}
	\text{Disc}[\text{Li}_n(z)] = \int\limits_1^z \text{Disc}[\text{Li}_{n-1}(t)]d\log(t) 
\end{align}
Which from the discontinuity of the logarithm tells us that 
\begin{align}
	\text{Disc}[\text{Li}_n(z)] = \frac{2\pi i}{(n-1)!}\log^{n-1}z
\end{align}
This is exactly $2\pi i$ times the function corresponding to the symbol we get from deleting the first entry of the symbol for the classical polylogarithm, the branch point being where the leading argument equals 0, in this case $z = 1$.  This is a general phenomenon.  Read from left to right, the symbol gives you the sequence of discontinuities of the function.  Additionally, the differential indicates that read from right to left, the symbol gives you the differential equations obeyed by the function.\\
\indent Finally we mention the integrability condition for symbols which is that for a symbol 
\begin{align*}
	S = \sum\limits_{I = (i_1, \dots, i_n)}c_{I} a_{i_1}\otimes \dots \otimes a_{i_n}
\end{align*}
We must have 
\begin{align*}
 \sum\limits_{I = (i_1, \dots, i_n)}c_{I} a_{i_1}\otimes \dots \otimes a_{i_{p-1}}\otimes a_{i_{p+2}} \otimes a_{i_n} d\log a_{i_{p}} \ \wedge d\log a_{i_{p+1}} = 0
\end{align*}
For every $p$, where we are replacing a symbol word by the $d$$\log$ of adjacent entries wedged times the symbol with the entries deleted.  This condition is non-trivial, as it is possible to write down symbols which do not obey the integrability condition and do not lift to a function.  We also note here that the symbol only fixes the piece of the function of highest transcendental weight.  It does not detect the terms of lower transcendental weight times transcendental numbers that still sum up to transcendental weight of the full function.   This is intuitive from the recursive definition for the symbol used here, as $d \log$ of a constant vanishes. For our purposes, these terms must be fixed by  boundary conditions. \\
\indent In order to illustrate the lucidity offered by the symbol, we will now prove the Abel dilogarithm identity.  First consider 
\begin{align*}
	\mathcal{S}\left(\li_2\left(\frac{wz}{(1-w)(1-z)} \right)\right)  =  -\frac{1-w-z}{(1-w)(1-z)}\otimes \frac{wz}{(1-w)(1-z)}
\end{align*}
We can use the linearity properties of the symbol to separate this into 
\begin{align*}
	\mathcal{S}\left(\li_2\left(\frac{wz}{(1-w)(1-z)} \right)\right)  = &-\frac{1-w-z}{1-z}\otimes \frac{w}{1-z}-\frac{1-z-w}{1-w}\otimes \frac{z}{1-w}-(1-w)\otimes w\\
	&-(1-z)\otimes z-(1-w)\otimes (1-z)-\otimes(1-z)\otimes(1-w)
\end{align*}
We clearly recognize these pieces as the symbol that gives us the equality
\begin{align*}
	\li_2\left(\frac{wz}{(1-w)(1-z)} \right) = \li_2\left(\frac{z}{1-w} \right)+\li_2\left(\frac{w}{1-z} \right)-\li_2(w)-\li_2(z)-\li_1(w)\li_1(z)
\end{align*}
There is also the possibility of functions of lower transcendentality times transcendental numbers.  At weight two we would only get a constant, so it remains to fix the constant.  Plugging in $w = 0$ we see that both sides vanish and so the equality holds exactly where it is well defined i.e. $w, z \notin (1, \infty)$.
\section{Statement of the Recursive Rule}\label{rule}
A graph $\mathcal{G}$ corresponds to some contribution to the perturbative wave function, which is some function $F_\mathcal{G}$ of the final state energies (up to the overall constant).  The discontinuities of this function occur when the positive sums of energies and cut edges corresponding to subgraphs of $\mathcal{G}$ become negative.  We denote such an energy corresponding to some subgraph $\mathfrak{g}$ as $E_\mathfrak{g}^{tot} = \sum\limits_{v \in \mathcal{V}} x_v +\sum\limits_{e_c\in \mathcal{E}^c} y_{e_c}$ where $e_c$ denotes an edge cut by circling the subgraph.  We diagrammatically denote the discontinuity of $F_\mathcal{G}$ corresponding to the subgraph as 
\begin{align}
	\text{Disc}_\mathfrak{g}[F_\mathcal{G}] = \begin{tikzpicture}[baseline=(c)]
	\begin{feynman}[inline=(c)]
	\vertex[] (c) at (0, 0) {};
	\vertex[] (upright) at (1, 1) {};	
	\vertex[] (upleft) at (-1, 1) {};	
	\vertex[] (down) at (0, -1) {};
	\vertex[] (centerupleft) at (-.25, .25) {};
	\vertex[] (centerdownright) at (.25, -.25) {};
	\vertex[] (centerupright) at (.25, .25) {};
	\vertex[] (centerdownleft) at (-.25, -.25) {};
	\vertex[] (centerright) at (.4, 0) {};
	\vertex[] (centerleft) at (-.4, 0) {};
	\diagram*{
	(upright) -- (centerupleft), 
	(upright) -- (centerdownright),
	(upleft) -- (centerupright), 
	(upleft) -- (centerdownleft),
	(down) -- (centerleft),
	(down) -- (centerright)
	};
	\draw[fill=white]  (0, 0) circle (.48 cm) node {$\mathfrak{g}$} ;
	\draw[fill=white]  (1, 1) circle (.48cm) node {$\overline{\mathfrak{g}}_i$} ;
	\draw[fill=white]  (-1, 1) circle (.48 cm) node {$\overline{\mathfrak{g}}_1$} ;
	\draw[fill=white]  (0, -1.35) circle (.48 cm) node {$\overline{\mathfrak{g}}_j$} ;
	\draw [dotted ,domain= 80:100] plot ({1.2*cos(\x)}, {1.2*sin(\x)});
	 \draw [dotted , domain=0:-30] plot ({1.2*cos(\x)}, {1.2*sin(\x)});
	  \draw [dotted , domain=210:180] plot ({1.2*cos(\x)}, {1.2*sin(\x)});
	  \draw [dotted , domain=308:324] plot ({-1+.78*cos(\x)}, {1+.78*sin(\x)});
	   \draw [dotted , domain=218:234] plot ({1+.78*cos(\x)}, {1+.78*sin(\x)});
	   \draw [dotted , domain=83:98] plot ({.78*cos(\x)}, {-1.39+.78*sin(\x)});
	   \draw [dashed] (0, 0) circle (.55 cm) {};
	\end{feynman}
\end{tikzpicture}
\end{align}
With the dashed circle enclosing the subgraph corresponding to the discontinuity we are taking.   Each complementary subgraph $\overline{\mathfrak{g}}_i$ has some set of vertices $\mathcal{V}^c_i$ which has cut edges, the edges which connect to $\mathfrak{g}$, ending on them.  We additionally denote this set of cut edges connecting to the complementary subgraph as $\mathcal{E}_i^c$.  We will demonstrate that the discontinuity obeys the equality 
\begin{align}
	\text{Disc}_\mathfrak{g}[F_\mathcal{G}(\{X_v\}, \{Y_e\})] = \text{Disc}_\mathfrak{g}[F_{\mathfrak{g}} (\{X_v\}, \{ X_{v_c}+Y_{e_c}\},\{Y_e\})]\times \nonumber \\ \prod\limits_i \sum\limits_{\{\sigma_{e_c}\}}(-1)^{n_+}F_{\overline{\mathfrak{g}}_i}(\{X_v\}, \{ X_{v_c}+\sigma_{e_c}Y_{e_c}\}, \{Y_e\})
\end{align}  
The set $\{X_{v_c} \}$ for any subgraph is the set of vertex energies corresponding to vertices with cut edges ending on them and the $Y_{e_c}$ are the corresponding edge energies.  The $\sigma_{e_c}$'s can take on values $\pm 1$ and the sum is over all such combinations, and $n_+$ counts the number of positive ones in $\{\sigma_{e_c}\}$, giving the appropriate sign.   Because the discontinuities are produced by poles in the energy integrand, the sum of all the discontinuities vanishes by the residue theorem.  The consequence of this at the level of the full function is that the total energy discontinuity is minus the sum of all other discontinuities.  This fact and the previous equation combine to produce a recursive rule for the symbol of any wave function contribution. Multiplication at the level of the integrated functions becomes a shuffle product at the level of the symbol and we tac on the appropriate energy sum as the leading symbol entry corresponding to the discontinuity.  That is,
\begin{align}
	\mathcal{S}(F_\mathcal{G}) = \sum\limits_{\mathfrak{g}}\frac{E_\mathfrak{g}^{tot}}{\left(\sum\limits_v X_v\right)}\otimes \mathcal{S}(\text{Disc}_\mathfrak{g}[F_{\mathfrak{g}} (\{X_v\}, \{ X_{v_c}+Y_{e_c}\},\{Y_e\})]) \nonumber\\
	  \bigshuffle_{i \textcolor{white}{oe}} \sum\limits_{\{\sigma_{e_c}\}}\left(-1 \right)^{n_+}\mathcal{S}(F_{\overline{\mathfrak{g}}_i}(\{X_v\}, \{ X_{v_c}+\sigma_{e_c}Y_{e_c}\}, \{Y_e\}))
\end{align}
  The sum over the signs of cut edges is a sum over possibilities of energy flowing into or out of the vertex, i.e. the propagation of an in or an out state.  Already we see that the discontinuity structure is the avatar of unitarity, providing a consistency condition for the factorization of discontinuities in terms of purely boundary data which we may then imbue with a natural bulk interpretation.  Moreover, this rule reveals that the symbol is the object that can be systematically and uniquely built up from its most elementary building block, the three-particle symbol: $X$, and consistent factorization of the discontinuities. We illustrate the simplicity and practicality of this recursive rule in Section \ref{examples}.
\subsection{Time Integral Derivation}
   Since the discontinuities are produced by poles in the integrand, it suffices to show that the energy integrand contributing to the perturbative wave function $\psi_n$ factorizes appropriately on the residue of each of its poles.  The poles of $f_\mathcal{G}$ correspond to the possible subgraphs of the diagram and are equal to the sum of the vertex energies plus the energies of edges running out of the subgraph.  We are interested in the residue as $E_\mathfrak{g}^{tot}$ goes to zero.  This pole is generated by the integration 
  \begin{align*}
  	 f_\mathcal{G} = \hspace{2mm} \dots\int\limits_{-\infty}^{\eta_c} d\eta_l e^{iE_{\mathfrak{g}}^{tot}\eta_l} 
  \end{align*}
  where $\eta_l$ is the latest time within $\mathfrak{g}$ for each contribution and $\eta_c$ is some cutoff time.  The dots are the later integrations to be carried out.  Such an integration only occurs when every vertex in $\mathfrak{g}$ is in the past of all vertices outside $\mathfrak{g}$ and as $E_{\mathfrak{g}}^{tot}$ goes to zero the integral is dominated by the region where $\eta_l \to -\infty$.  As such, when localized to the residue at this pole, all $\theta$ functions in the propagators inconsistent with times contained in the subgraph being in the past of vertices outside it lose support in this region.  We can organize the propagators suggestively as
   \begin{align}
   	f_{\mathcal{G}} = \left(\prod\limits_{e}2Y_e\right)\prod\limits_{\overline{\mathfrak{g}}_i}\left(\int\limits_{-\infty}^0 \prod\limits_{v \in \mathcal{V}_i}d\eta_v e^{ix_v\eta_v}\prod\limits_{e \in \mathcal{E}_i}G(\eta_{v_e}, \eta_{v_e'}, y_e) \right)\times & \nonumber \\ \int\limits_{-\infty}^0\prod\limits_{e_c \in \mathcal{E}_c}G_c(\eta_{e_c}^{out}, \eta_{e_c}^{in}, y_{e_c})&\left( \prod\limits_{v \in \mathcal{V}_\mathfrak{g}}d\eta_v e^{ix_v\eta_v}\prod\limits_{e \in \mathcal{E}_\mathfrak{g}}G(\eta_{v_e}, \eta_{v_e'}, y_e) \right)
   \end{align}
   Thus far we have merely expanded $f_\mathcal{G}$, suggestively organizing terms.  
   The first piece is a product over all disconnected subgraphs, with all the vertices and edges internal to the graph contained in the parentheses.  The right parenthetical contains the same but for the subgraph corresponding to the residue we are taking.  Between the two parentheticals we have the product over cut edges $e_c$ leading out from the subgraph $\mathfrak{g}$.  Now we localize to the $E_{\mathfrak{g}}^{tot}$ poles. Because $\theta(\eta_{e_c}^{in}-\eta_{e_c}^{out})$ has no support and the other $\theta$ function is satisfied we replace the propagator with the ``cut" propagator
   \begin{align}
   	G_c(\eta_{e_c}^{out}, \eta_{e_c}^{in}, y_{e_c}) = \frac{1}{2y_{e_c}} \left(e^{-iy_{e_c}\eta_{e_c}^{out}} -e^{iy_{e_c}\eta_{e_c}^{out}}\right)e^{iy_{e_c}\eta_{e_c}^{in}}
   \end{align}
  Where we omit the implicit theta function in the first term since it is already guaranteed when we are at the pole.  It is worth commenting on an exception to this which is possible with loops, where an edge leads out and back into the subgraph so that $\eta_{e_c}^{in} \equiv \eta_{e_c}^{out}$.  In this case, neither $\theta$ function has support, and the cut propagator only has the last term.  This is explicitly applied to computing the one-loop symbol in Section \ref{examples}.\\ 
  \indent We can now take the residue.  It is known that the residue of the total energy pole gives the flat-space scattering amplitude \cite{cosmocollider, cosmobootstrap, cosmopoly}, and this is the factor that remains from the subgraph corresponding to the residue.  Focusing on the external piece we are left with 
  \begin{align}
  \underset{E_{\mathfrak{g}}^{tot}\to 0}{\text{Res }}f_\mathfrak{g}^{ext} \propto \prod\limits_{\overline{\mathfrak{g}}_i}\left(\int\limits_{-\infty}^0 \prod\limits_{v \in \mathcal{V}_i/\mathcal{V}^c_i}d\eta_v e^{ix_v\eta_v}\prod\limits_{v \in \mathcal{V}^c_i}d\eta_v (e^{i(x_v-y_{e_c})\eta_v}-e^{i(x_v+y_{e_c})\eta_v})\prod\limits_{e \in \mathcal{E}_i}G(\eta_{v_e}, \eta_{v_e'}, y_e) \right)
  \end{align}  
  where $\mathcal{V}_i^c$ are the vertices in subgraph $\overline{\mathfrak{g}}_i$ that have a cut edge ending on them.  The above is merely the the sum over combinations of absorbing positive or negative edge energy for each vertex with a cut edge ending on it, with a product then taken over subgraphs.  All the time-integrals between subgraphs are totally decoupled and so the full expression for the residue is 
  \begin{align}
  	\underset{E_{\mathfrak{g}}^{tot}\to 0}{\text{Res }} f_\mathcal{G} = 2\pi i\mathcal{A}[\mathfrak{g}]\times \prod\limits_{\overline{\mathfrak{g}}_i}\sum_{\{\sigma_{e_c} \}}\left(-1\right)^{n_+} f_{\overline{\mathfrak{g}}_i}(\{x_v\}, \{x_{v_c}+\sigma_{e_c} y_{e_c} \}, \{y_e\})
  \end{align}
  Where $\mathcal{A}[g]$ denotes the flat space scattering amplitude corresponding to the subgraph $\mathfrak{g}$ and
  we have distinguished between vertices $v_c$ and $v$ which do and do not have cut edges ending on them respectively.  The $\sigma_{e_c}$'s are plus or minus one and $n_+$ enforces the appropriate sign depending on whether there is an even or odd number of edges absorbing positive energy.  Because the residue of the integrand is fully factorized, with no energy dependence mixing between subgraphs, the energy integral is just a product over the integrals for each subgraph and so this factorization trivially lifts to the level of the full wave function contribution, with the residue at the pole now corresponding to the discontinuity across a branch cut.
   \subsection{From the Cosmological Polytope}
   We can also understand the origin of the recursive rule from the perspective of the cosmological polytope, where the ingredients are factorization and triangulation.  We will illustrate the rule utilizing two discontinuities of the three-site chain as the facets of corresponding polytope can be visualized in three dimensions.  We use the formalism laid out in the latter half of \cite{cosmopoly}. The cosmological polytope corresponding to this graph is the ``double-square pyramid" which has six vertices and lives in $\mathbb{P}^4$.  We first examine the discontinuity associated with $X_1+Y_1$.  The corresponding facet of the cosmological polytope has the five vertices indicated by the marking 
   \begin{align*}
   	 \begin{tikzpicture}
     \begin{feynman}
     	\vertex (v1);
     	\vertex[right=1cm] (v2);
     	\vertex[right=2cm of v1] (v3);
     	\vertex[left=.05cm] (v4) {\tiny $x_1$} ;
 		\vertex[right=1cm, below=.05cm of v2] (v5) {\tiny $x_2$};
 		\vertex[right=2.05cm]  (v6) {\tiny $x_3$};
     	\diagram*{
     	(v1) -- (v2) -- (v3)
     	};
     	\draw[fill=black] (v1) circle(.5mm);
\draw[fill=black] (v2) circle(.5mm);
\draw[fill=black] (v3) circle(.5mm);
\draw[dashed] (0, 0) circle (1.2mm);
\draw[thick] (.5, 0) circle (.7mm) ;
\draw[thick] (.8, 0) circle (.7mm) ;
\draw[thick] (1.5, 0) circle (.7mm) ;
\draw[thick] (1.8, 0) circle (.7mm) ;
\draw[thick] (1.2, 0) circle (.7mm) ;
\draw[thick] (.23, 0) node {$\times$};
     \end{feynman}
 \end{tikzpicture}
   \end{align*}
   We can visualize this facet in three-dimensional space as it's constructed out of intersecting triangles
   
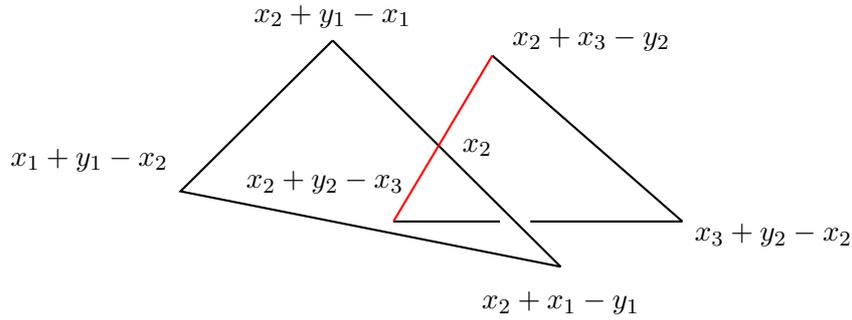
\begin{align*}
   	\begin{tikzpicture}[scale = 2.0, baseline=10mm]
	\node (x2) at (0, 0) [label={[xshift = -.1cm, yshift=0.2cm] $x_2$}] {};
	\coordinate (vL1) at (-1, 1) ;
	\coordinate (vL2) at (-2, 0);
	\node (vL2label) at (-2, 0) [label={[xshift = -1.2cm, yshift=0cm] $x_1+y_1-x_2$}] {};
	\coordinate (vL30) at (-.5, .5);
	\node (vL3label) at (-.5, .5) [label={[xshift = -1cm, yshift=.9cm] $x_2+y_1-x_1$}] {};
	\coordinate (vL31) at (-.3, .3);
	\coordinate (vL32) at (.5, -.5);
	\node (vL3label) at (.5, -.5) [label={[xshift = 0cm, yshift=-.9cm] $x_2+x_1-y_1$}] {};
	\coordinate(vR10) at (-.6, -.2);
	\node (vR10label) at (-.6, -.2) [label={[xshift = -.9cm, yshift=.1cm] $x_2+y_2-x_3$}] {};
	\coordinate(vR11) at (.1, -.2);
	\coordinate(vR12) at (.3, -.2);
	\coordinate(vR2) at (1.3, -.2);
	\node (vR2label) at (1.3, -.2) [label={[xshift = 1.2cm, yshift=-.6cm] $x_3+y_2-x_2$}] {};
	\coordinate(vR3) at (.05, .9);
	\node (vR3label) at (.05, .9) [label={[xshift = 1.3cm, yshift=-.2cm] $x_2+x_3-y_2$}] {};
	\begin{scope}[thick]
	\draw (vL1) -- (vL2) -- (vL32);
	\draw (vL32) -- (vL1);
   \draw (vR10) -- (vR11);
   \draw (vR12) -- (vR2);
   \draw (vR2) -- (vR3);
   \draw[color=red] (vR3) -- (vR10);
	\end{scope}
\end{tikzpicture}
   \end{align*} 
   The vertex opposite the red line at $x_1+y_1-x_2$ moves off in a fourth dimension and is the vertex not on this facet of the cosmological polytope.  This red line also provides the natural triangulating hyperplane of the facet needed to compute the canonical form.  The two tetrahedra that result from triangulating the pyramid give the canonical forms with $x_2$ absorbing positive and negative edge energies.
   
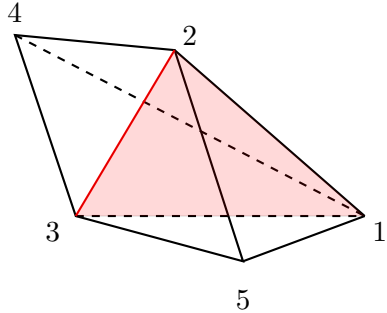
\begin{align*}
   	\begin{tikzpicture}[scale = 2, baseline=5mm]
	\coordinate (x2) at (0, 0);
	\coordinate (vL1) at (-1, 1);
	\coordinate (vL3) at (.5, -.5);
	\node (vL3label) at (-.5, .5) [label={[xshift = -1cm, yshift=.9cm] $\text{4}$}] {};
	\node (vL3label) at (.5, -.5) [label={[xshift = 0cm, yshift=-.9cm] $\text{5}$}] {};
	\coordinate(vR1) at (-.6, -.2);
	\node (vR10label) at (-.6, -.2) [label={[xshift = -.3cm, yshift=-.6cm] $\text{3}$}] {};
	\coordinate(vR2) at (1.3, -.2);
		\node (vR2label) at (1.3, -.2) [label={[xshift = .2cm, yshift=-.6cm] $\text{1}$}] {};
	\coordinate(vR3) at (.05, .9);
	\node (vR3label) at (.05, .9) [label={[xshift = .2cm, yshift=-.2cm] $\text{2}$}] {};
	\begin{scope}[thick]
	\draw (vL1) -- (vR3) -- (vL3) -- (vR1) -- cycle;
	\draw[dashed] (vL1) -- (vR2);
	\draw[dashed] (vR1) -- (vR2);
	\draw (vR3) -- (vR2);
	\draw (vL3) -- (vR2);
	\draw[color=red] (vR3) -- (vR1);
	\end{scope}
	\draw[fill=red, opacity =0.15] (vR3) --(vR1) -- (vR2);
\end{tikzpicture}
   \end{align*} 
   It is clear that the red shaded triangle comprising the intersection of a hyperplane with the pyramid is the polytope in $\mathbb{P}^2$ corresponding to the right triangle, which is the cosmological polytope corresponding to the Feynman diagram made of vertices $x_2$ and $x_3$ connected by edge $y_2$.  Therefore the canonical form on this two dimensional subspace is 
   \begin{align}
   	\Omega = \frac{\braket{Xd^2X}\braket{123}^2}{\braket{X12}\braket{X23}\braket{X 31}}
   \end{align}
   and its wave function contribution is that of the two-site chain.
   In this triangulation, the canonical form for the pyramid is the difference of the canonical forms on two tetrahedra separated by the hyperplane.
   We see this is 
   \begin{align}
   	\Omega =\frac{1}{\braket{X123}}\left(\frac{\braket{Xd^3X}\braket{1234}^3}{\braket{X234}\braket{X341}\braket{X412}}-\frac{\braket{Xd^3X}\braket{1325}^3}{\braket{X325}\braket{X251}\braket{X513}} \right)
   \end{align}  
  Vertices 4 and 5 of the pyramid correspond to the two circles left of the two-site subgraph and are the only vertices lifted into the $y_1$ direction. We see that  
  \begin{align}
  	\Omega =\frac{1}{2y_1}\left( \frac{\braket{Xd^3X}\braket{1234}^3}{(\braket{X23}-2y_1)(\braket{X31}-y_1)(\braket{X12}-2y_1)}-\frac{\braket{Xd^3X}\braket{1325}^3}{(\braket{X23}+2y_1)(\braket{X12}+2y_1)(\braket{X31}+2y_1)}\right)
  \end{align}
 The piece corresponding to the wave function contribution is then 
 \begin{align}
 	\underset{x_1+y_1 \to 0 }{\text{Res}} \psi  = \frac{2\pi i}{2y_1}\left(\frac{1}{(x_3+y_2)(x_2-y_1+y_2)(x_3+x_2-y_1)} -\frac{1}{(x_3+y_2)(x_2+y_1+y_2)(x_3+x_2+y_1)}  \right)
 \end{align}
 consistent with the recursive rule.  This is a general mechanism; the cosmological polytope corresponding to the subgraph is a triangulating hyperplane, separating polytopes whose canonical forms correspond to absorbing positive and negative edge energy.\\
 \indent The other aspect to understand from the polytope perspective is the factorization of disconnected subgraphs.  This can be illustrated with the following discontinuity 
 	   \begin{align}
   	 \begin{tikzpicture}
     \begin{feynman}
     	\vertex (v1);
     	\vertex[right=1cm] (v2);
     	\vertex[right=2cm of v1] (v3);
     	\vertex[left=.05cm] (v4) {\tiny $x_1$} ;
 		\vertex[right=1cm, below=.05cm of v2] (v5) {\tiny $x_2$};
 		\vertex[right=2.05cm]  (v6) {\tiny $x_3$};
     	\diagram*{
     	(v1) -- (v2) -- (v3)
     	};
     	\draw[fill=black] (v1) circle(.5mm);
\draw[fill=black] (v2) circle(.5mm);
\draw[fill=black] (v3) circle(.5mm);
\draw[dashed] (1, 0) circle (1.2mm);
\draw[thick] (.5, 0) circle (.7mm) ;
\draw[thick] (.75, 0) node {$\times$} ;
\draw[thick] (1.5, 0) circle (.7mm) ;
\draw[thick] (1.8, 0) circle (.7mm) ;
\draw[thick] (1.25, 0) node {$\times$} ;
\draw[thick] (.23, 0) circle (.7mm) ;
     \end{feynman}
 \end{tikzpicture}
   \end{align}
   Which is the discontinuity associated with $X_2+Y_1+Y_2$.
This facet of the cosmological polytope is a tetrahedron.
\begin{align}
\begin{tikzpicture}[scale=2]
     \coordinate (y1r) at (.3, .2);
	\coordinate (y1l) at (-1.2, .5);
	\coordinate (y2r) at (.8, 1.3);
	\coordinate (y2l) at (-.3, 2);
	\node (y2llabel) at (-.4, 2.2) {$x_3-y_2$};
	\node (y1llabel) at (-1.7, .5) {$x_1-y_1$};
	\node (y1rlabel) at (.8, .2) {$x_1+y_1$};
	\node (y2rlabel) at (1.2, 1.4) {$x_3+y_2$};
	\begin{scope}[thick]
		\draw (y1r) -- (y2r);
		\draw (y1l) -- (y2l);
		\draw (y1r) -- (y2l);
		\draw[blue] (y1r) -- (y1l);
		\draw[red] (y2r) -- (y2l);
		\draw[dashed] (y1l) -- (y2r);
	\end{scope}
\end{tikzpicture}
\end{align}
The red and blue lines are subspaces in $\mathbb{P}^1$.  We can call these projective subspaces $X_1$ and $X_2$.  The full tetrahedron in $\mathbb{P}^3$ is the Cartesian product of these subspaces.  It can be understood as made from combining the subspaces with projective weights $ \tau_1 X_1+\tau_2 X_2$ where the $\tau$'s are positive.  Each subspace has one projective degree of freedom and the simplex of $\tau$'s is the third projective degree of freedom which we can describe with one projective coordinate $\tau$.  The canonical form on the tetrahedron is then the three-form 
 \begin{align}
 	\Omega_X = \Omega_{X_1}\wedge \Omega_{\tau}\wedge \Omega_{X_2}
 \end{align}
We will show that this works explicitly.  We can call the blue subspace $X_1$ and the red $X_2$.  The vertices on $X_1$ are $\{x_1+y_1, x_1-y_1\}$ which we call $v_1$ and $v_2$. $X_1$ is 
\begin{align}
	X_1 = \alpha_1 v_1+\alpha_2 v_2
\end{align}
The canonical form is then 
\begin{align}
	\Omega_{X_1} = d\log\left(\frac{\alpha_1}{\alpha_2} \right)
\end{align}
with poles at 0 and $\infty$ in the projective parameter $\alpha = \frac{\alpha_1}{\alpha_2}$.  In homogenous coordinates we have 
\begin{align}
	X_1 = \alpha_1\begin{pmatrix}
		1 \\ 1
	\end{pmatrix}+\alpha_2 \begin{pmatrix}
		1 \\ -1
	\end{pmatrix} = \begin{pmatrix}
		x_1 \\ y_1
	\end{pmatrix}
\end{align} 
and so we get the appropriate form 
\begin{align}
	\Omega_{X_1} = d\log\left(\frac{x_1-y_1}{x_1+y_1} \right)
\end{align}
This is just the usual canonical form for an interval in $\mathbb{P}^1$.  The same holds for $X_2$ and we call the parameters $\beta_1, \beta_2$.  Now, with the combined space we simply have 
\begin{align}
	X = \tau_1 v_1+\tau_1 \alpha v_2 + \tau_2 v_3+\tau_2\beta v_4
\end{align}
And we then say the canonical form is 
\begin{align}
	\Omega_X = d\log \alpha\wedge d\log \beta  \wedge  d \log\left(\frac{\tau_1}{\tau_2} \right)
\end{align}
Now to extract the form in the homogenous coordinates we again proceed 
\begin{align}
	X = \tau_1 \begin{pmatrix}
		1 \\ 1 \\ 0 \\ 0
	\end{pmatrix}+ \tau_1\alpha \begin{pmatrix}
		1 \\ -1 \\ 0 \\ 0
	\end{pmatrix}+ \tau_2 \begin{pmatrix}
		0 \\ 0 \\ 1 \\ 1
	\end{pmatrix}+ \tau_2 \beta \begin{pmatrix}
		0 \\ 0 \\ 1 \\ -1
	\end{pmatrix}
\end{align}
and from these the form is given by
\begin{align}
	\Omega_X = d\log\left(\frac{x_1-y_1}{x_1+y_1} \right)\wedge d\log\left(\frac{x_3-y_2}{x_3+y_2} \right)\wedge d\log\left(\frac{x_1+y_1}{x_3+y_2}\right)
\end{align}
Which does give the correct residue for the wave function contribution.  If we added another subgraph, we would be working with the cartesian product of this original polytope and the polytope of the subgraph and the same would hold, and so the factorization can be seen inductively.
It is worth commenting that different gauge-fixing choices for the projective weights would result in forms that do not in general have the same differentials.  This is merely an artifact of having to make a choice, but the contribution to the perturbative wave function is the same no matter choice we make.  We see that the essential ingredients for symbol recursion are elegantly encoded in the geometry of the cosmological polytope.
\section{Some Examples}\label{examples}
\subsection{Two-Site Chain}
The symbol for the two-site chain, corresponding to the four-point correlator, is readily acquired with the recursive rule. The graphical rule states
\begin{align}
	  \begin{tikzpicture}
     \begin{feynman}
     	\vertex (v1);
     	\vertex[right=1cm] (v2);
     	\vertex[left=.05cm] (v3) {\tiny $x_1$} ;
 		\vertex[right=1.05cm] (v4) {\tiny $x_2$};
 		\vertex[left=.15cm of v1] (v5);
 		\vertex[right=.15cm] (v6);
     	\diagram*{
     	(v1) -- (v2),
     	(v5) -- [half left, dashed] (v6) -- [half left, dashed] (v5)
     	};
     	\draw[fill=black] (v1) circle(.5mm);
\draw[fill=black] (v2) circle(.5mm);
     \end{feynman}
 \end{tikzpicture} =  \begin{tikzpicture}
     \begin{feynman}
     	\vertex (v1);
     	\vertex[right=.95cm] (v2);
     	\vertex[left=-.01cm] (v3) {\tiny $-$} ;
 		\vertex[right=1.0cm] (v4) {\tiny $x_2$};
     	\diagram*{
     	(v1) -- (v2)
     	};
\draw[fill=black] (v2) circle(.5mm);
     \end{feynman}
 \end{tikzpicture}-\begin{tikzpicture}
     \begin{feynman}
     	\vertex (v1);
     	\vertex[right=.95cm] (v2);
     	\vertex[left=.01cm] (v3) {\tiny $+$} ;
 		\vertex[right=1.0cm] (v4) {\tiny $x_2$};
     	\diagram*{
     	(v1) -- (v2)
     	};
\draw[fill=black] (v2) circle(.5mm);
     \end{feynman}
 \end{tikzpicture}
\end{align}
The same holds for the other vertex and so the symbol is immediately determined
\begin{align}
	\mathcal{S} = \frac{X_1+Y}{X_1+X_2}\otimes  \frac{X_2-Y}{X_2+Y}+\frac{X_2+Y}{X_1+X_2}\otimes  \frac{X_1-Y}{X_1+Y}
\end{align}
The symbol is naturally expressed in the following variables
\begin{align}
	u = \frac{X_1-Y}{X_1+X_2}\\
	v = \frac{X_2-Y}{X_1+X_2}
\end{align}
and it takes the form 
\begin{align}
	\mathcal{S} = (1-u)\otimes \frac{u}{1-v}+(1-v)\otimes\frac{v}{1-u}
\end{align}
\subsection{Two-Site Loop}
Here, the graphical rule on the vertices says 
\begin{align}
	\begin{tikzpicture}[baseline=(v4)]
 	\begin{feynman}[inline=(v4)]
 		\vertex (v1) ;
 		\vertex[right=1cm] (v2);
 		\vertex[left=.05cm] (v3) {\tiny $x_1$} ;
 		\vertex[right=1.05cm] (v4) {\tiny $x_2$};
 		\vertex[left=.15cm of v1] (v5);
 		\vertex[right=.15cm] (v6);
 		\diagram*{
 		(v1) -- [half left, edge label =\tiny \( y_a\)] (v2) -- [half left, edge label = \tiny \( y_b\)] (v1),
 		(v5) -- [half left, dashed] (v6) -- [half left, dashed] (v5)
 		};
 		\draw[fill=black] (v1) circle(.5mm);
\draw[fill=black] (v2) circle(.5mm);
 	\end{feynman}
 \end{tikzpicture}
=
  \hspace{7mm}\begin{tikzpicture}[baseline=(v5)]
     \begin{feynman}[inline=(v5)]
     	\vertex (v1);
     	\vertex[above=1cm] (v2);
     	\vertex[above=2cm of v1] (v3) {\tiny $+$};
     	\vertex[below=.05cm] (v4) {\tiny $+$} ;
 		\vertex[above=1cm, right=.05cm of v2] (v5) {\tiny $x_2$};
 		\vertex[above=2.05cm]  (v6) ;
     	\diagram*{
     	(v1) -- (v2) -- (v3)
     	};
\draw[fill=black] (v2) circle(.5mm);
\end{feynman}
\end{tikzpicture}\hspace{2mm} - \hspace{2mm} \begin{tikzpicture}[baseline=(v5)]
     \begin{feynman}[inline=(v5)]
     	\vertex (v1);
     	\vertex[above=1cm] (v2);
     	\vertex[above=2cm of v1] (v3) {\tiny $-$};
     	\vertex[below=.05cm] (v4) {\tiny $+$} ;
 		\vertex[above=1cm, right=.05cm of v2] (v5) {\tiny $x_2$};
 		\vertex[above=2.05cm]  (v6) ;
     	\diagram*{
     	(v1) -- (v2) -- (v3)
     	};
\draw[fill=black] (v2) circle(.5mm);
\end{feynman}
\end{tikzpicture}\hspace{2mm} - \hspace{2mm}  \begin{tikzpicture}[baseline=(v5)]
     \begin{feynman}[inline=(v5)]
     	\vertex (v1);
     	\vertex[above=1cm] (v2);
     	\vertex[above=2cm of v1] (v3) {\tiny $+$};
     	\vertex[below=.05cm] (v4) {\tiny $-$} ;
 		\vertex[above=1cm, right=.05cm of v2] (v5) {\tiny $x_2$};
 		\vertex[above=2.05cm]  (v6) ;
     	\diagram*{
     	(v1) -- (v2) -- (v3)
     	};
\draw[fill=black] (v2) circle(.5mm);
\end{feynman}
\end{tikzpicture}\hspace{2mm} + \hspace{2mm}  \begin{tikzpicture}[baseline=(v5)]
     \begin{feynman}[inline=(v5)]
     	\vertex (v1);
     	\vertex[above=1cm] (v2);
     	\vertex[above=2cm of v1] (v3) {\tiny $-$};
     	\vertex[below=.05cm] (v4) {\tiny $-$} ;
 		\vertex[above=1cm, right=.05cm of v2] (v5) {\tiny $x_2$};
 		\vertex[above=2.05cm]  (v6) ;
     	\diagram*{
     	(v1) -- (v2) -- (v3)
     	};
\draw[fill=black] (v2) circle(.5mm);
\end{feynman}
\end{tikzpicture}
\end{align}
The same holds for the second vertex and so the vertices  contribute 
\begin{align}
	\frac{X_1+Y_a+Y_b}{X_1+X_2} \otimes\left[(X_2+Y_a+Y_b)-(X_2-Y_a+Y_b) -(X_2+Y_a-Y_b)+(X_2-Y_a-Y_b) \right]\nonumber\\
	\frac{X_2+Y_a+Y_b}{X_1+X_2} \otimes\left[(X_1+Y_a+Y_b)-(X_1-Y_a+Y_b) -(X_1+Y_a-Y_b)+(X_1-Y_a-Y_b) \right]\nonumber
\end{align}
where we summed over all combinations of edge energy signs inside the brackets.  In order to get the contribution for the larger subgraphs, we use the fact that the discontinuity of the subgraph is minus the sum of the discontinuities of $\textit{its}$ subgraphs, with positive energy absorbed from cut edges.  
Graphically this means we have the following
\begin{align}
	\begin{tikzpicture}[baseline=(leftout)]
 	\begin{feynman}[inline=(leftout)]
 		\vertex (v1) ;
 		\vertex[right=1cm] (v2);
 		\vertex[left=.05cm] (v3) {\tiny $x_1$} ;
 		\vertex[right=1.05cm] (v4) {\tiny $x_2$};
 		\vertex[right=.15cm] (leftin);
 		\vertex[left=.15cm] (leftout);
 		\vertex[right=.85cm] (rightin);
 		\vertex[right=1.15cm] (rightout);
 		\diagram*{
 		(v1) -- [half left, edge label =\tiny \( y_a\)] (v2) -- [half left, edge label = \tiny \( y_b\)] (v1),
 		(leftin) -- [half left, dashed] (leftout) -- [half left, dashed] (rightout) -- [half left, dashed] (rightin) -- [half right, dashed] (leftin)
 		};
 		\draw[fill=black] (v1) circle(.5mm);
\draw[fill=black] (v2) circle(.5mm);
\draw[fill=black] (v2) circle(.5mm);
 	\end{feynman}
 \end{tikzpicture} = \hspace{7mm}-\begin{tikzpicture}[baseline=(v5)]
     \begin{feynman}[inline=(v5)]
     	\vertex (v1);
     	\vertex[above=1cm] (v2);
     	\vertex[above=2cm of v1] (v3) {\tiny $-$};
     	\vertex[below=.05cm] (v4) {\tiny $+$} ;
 		\vertex[above=1cm, right=.05cm of v2] (v5) {\tiny $x_1$};
 		\vertex[above=2.05cm]  (v6) ;
     	\diagram*{
     	(v1) -- (v2) -- (v3)
     	};
\draw[fill=black] (v2) circle(.5mm);
\end{feynman}
\end{tikzpicture}\hspace{2mm} + \hspace{2mm} \begin{tikzpicture}[baseline=(v5)]
     \begin{feynman}[inline=(v5)]
     	\vertex (v1);
     	\vertex[above=1cm] (v2);
     	\vertex[above=2cm of v1] (v3) {\tiny $+$};
     	\vertex[below=.05cm] (v4) {\tiny $+$} ;
 		\vertex[above=1cm, right=.05cm of v2] (v5) {\tiny $x_1$};
 		\vertex[above=2.05cm]  (v6) ;
     	\diagram*{
     	(v1) -- (v2) -- (v3)
     	};
\draw[fill=black] (v2) circle(.5mm);
\end{feynman}
\end{tikzpicture}\hspace{2mm} - \hspace{2mm}  \begin{tikzpicture}[baseline=(v5)]
     \begin{feynman}[inline=(v5)]
     	\vertex (v1);
     	\vertex[above=1cm] (v2);
     	\vertex[above=2cm of v1] (v3) {\tiny $-$};
     	\vertex[below=.05cm] (v4) {\tiny $+$} ;
 		\vertex[above=1cm, right=.05cm of v2] (v5) {\tiny $x_2$};
 		\vertex[above=2.05cm]  (v6) ;
     	\diagram*{
     	(v1) -- (v2) -- (v3)
     	};
\draw[fill=black] (v2) circle(.5mm);
\end{feynman}
\end{tikzpicture}\hspace{2mm} + \hspace{2mm}  \begin{tikzpicture}[baseline=(v5)]
     \begin{feynman}[inline=(v5)]
     	\vertex (v1);
     	\vertex[above=1cm] (v2);
     	\vertex[above=2cm of v1] (v3) {\tiny $+$};
     	\vertex[below=.05cm] (v4) {\tiny $+$} ;
 		\vertex[above=1cm, right=.05cm of v2] (v5) {\tiny $x_2$};
 		\vertex[above=2.05cm]  (v6) ;
     	\diagram*{
     	(v1) -- (v2) -- (v3)
     	};
\draw[fill=black] (v2) circle(.5mm);
\end{feynman}
\end{tikzpicture}
\end{align}
The contributions for the two U-shaped subgraphs are therefore
\begin{align}
	\frac{X_1+X_2+2Y_b}{X_1+X_2} \otimes\left[\frac{X_1+Y_b-Y_a}{X_1+Y_b+Y_a}+ \frac{X_2+Y_b-Y_a}{X_2+Y_b+Y_a}\right]\nonumber\\
	\frac{X_1+X_2+2Y_a}{X_1+X_2} \otimes\left[\frac{X_1+Y_a-Y_b}{X_1+Y_b+Y_a}+ \frac{X_2+Y_a-Y_b}{X_2+Y_b+Y_a}\right]\nonumber
\end{align}
If we add zero by both adding and subtracting each of $X_1+Y_a+Y_b\otimes X_2+Y_a+Y_b$ and $X_2+Y_a+Y_b\otimes X_1+Y_a+Y_b$, thereby contributing four new terms, we are motivated to introduce three sets of variables analogous to the two site 
\begin{align}
	u_1 &= \frac{X_1+Y_a-Y_b}{X_1+X_2+2Y_a}  \hspace{3mm} &u_2 &= \frac{X_1+Y_b-Y_a}{X_1+X_2+2Y_b}  \hspace{3mm} &u_3 &= \frac{X_1-Y_a-Y_b}{X_1+X_2} \nonumber\\
	v_1 &= \frac{X_2+Y_a-Y_b}{X_1+X_2+2Y_a}  \hspace{3mm} &v_2 &= \frac{X_2+Y_b-Y_a}{X_1+X_2+2Y_b}  \hspace{3mm} &v_3 &= \frac{X_2-Y_a-Y_b}{X_1+X_2}  \nonumber
\end{align}
The symbol is neatly expressed as 
\begin{align}
	\mathcal{S}_{\text{1-loop}}= &-(1-u_1)\otimes \frac{u_1}{1-v_1}-(1-v_1)\otimes\frac{v_1}{1-u_1}\nonumber\\
	&-(1-u_2)\otimes \frac{u_2}{1-v_2}-(1-v_2)\otimes\frac{v_2}{1-u_2}\nonumber\\
	&(1-u_3)\otimes \frac{u_3}{1-v_3}+(1-v_3)\otimes\frac{v_3}{1-u_3} 
\end{align}
We recognize this as three two-site chains, a fact that can also be seen by appropriately rearranging the time-ordered Feynman diagrams.
\subsection{Three-Site Chain}
The previous two examples illustrate the intuitive diagrammatic organization of symbol contributions.  The three-site chain is the first example which additionally demonstrates the efficiency of the recursive rule in computing the symbol.  We begin by computing the contributions from vertices
\begin{align}
	   \begin{tikzpicture}
     \begin{feynman}[inline=(v1)]
     	\vertex (v1);
     	\vertex[right=1cm] (v2);
     	\vertex[right=2cm of v1] (v3);
     	\vertex[left=.05cm] (v4) {\tiny $x_1$} ;
 		\vertex[right=1cm, below=.05cm of v2] (v5) {\tiny $x_2$};
 		\vertex[right=2.05cm]  (v6) {\tiny $x_3$};
 		\vertex[left=.15cm of v1] (v7);
 		\vertex[right=.15cm of v1] (v8);
     	\diagram*{
     	(v1) -- (v2) -- (v3),
     	(v7) -- [half left, dashed] (v8) -- [half left, dashed] (v7)
     	};
     	\draw[fill=black] (v1) circle(.5mm);
\draw[fill=black] (v2) circle(.5mm);
\draw[fill=black] (v3) circle(.5mm);
     \end{feynman}
 \end{tikzpicture} =    \begin{tikzpicture}
     \begin{feynman}[inline=(v1)]
     	\vertex (v1);
     	\vertex[right=1cm] (v2);
     	\vertex[right=2cm of v1] (v3);
     	\vertex[left=.05cm] (v4) {\tiny $-$} ;
 		\vertex[right=1cm, below=.05cm of v2] (v5) {\tiny $x_2$};
 		\vertex[right=2.05cm]  (v6) {\tiny $x_3$};
     	\diagram*{
     	(v1) -- (v2) -- (v3)
     	};
\draw[fill=black] (v2) circle(.5mm);
\draw[fill=black] (v3) circle(.5mm);
     \end{feynman}
 \end{tikzpicture}-
 \begin{tikzpicture}
     \begin{feynman}[inline=(v1)]
     	\vertex (v1);
     	\vertex[right=1cm] (v2);
     	\vertex[right=2cm of v1] (v3);
     	\vertex[left=.05cm] (v4) {\tiny $+$} ;
 		\vertex[right=1cm, below=.05cm of v2] (v5) {\tiny $x_2$};
 		\vertex[right=2.05cm]  (v6) {\tiny $x_3$};
     	\diagram*{
     	(v1) -- (v2) -- (v3)
     	};
\draw[fill=black] (v2) circle(.5mm);
\draw[fill=black] (v3) circle(.5mm);
     \end{feynman}
 \end{tikzpicture}
\end{align}
This is the difference of two-site chains where the absorbed edge energy is either positive or negative.  Therefore the contributions for the external vertices are
\begin{align*}
	\frac{X_1+Y_1}{X_1+X_2+X_3} \otimes \bigg[\frac{X_3+Y_2}{X_3+X_2-Y_1}\otimes  \frac{X_2-Y_1-Y_2}{X_2-Y_1+Y_2}+\frac{X_2-Y_1+Y_2}{X_3+X_2-Y_1}\otimes  \frac{X_3-Y_2}{X_3+Y_2}\\
	 -\frac{X_3+Y_2}{X_3+X_2+Y_1}\otimes  \frac{X_2+Y_1-Y_2}{X_2+Y_1+Y_2}-\frac{X_2+Y_1+Y_2}{X_3+X_2+Y_1}\otimes  \frac{X_3-Y_2}{X_3+Y_2}\bigg]\\\\
    \frac{X_3+Y_2}{X_1+X_2+X_3}\otimes \bigg [ \frac{X_1+Y_1}{X_1+X_2-Y_2}\otimes  \frac{X_2-Y_2-Y_1}{X_2-Y_2+Y_1}+\frac{X_2-Y_2+Y_1}{X_1+X_2-Y_1}\otimes  \frac{X_1-Y_1}{X_1+Y_1}\\
	 -\frac{X_1+Y_1}{X_1+X_2+Y_2}\otimes  \frac{X_2+Y_2-Y_1}{X_2+Y_2+Y_1}-\frac{X_2+Y_2+Y_1}{X_1+X_2+Y_2}\otimes  \frac{X_1-Y_1}{X_1+Y_1}\bigg]
\end{align*}
Next we look at the middle vertex. We have 
\begin{align}
      \begin{tikzpicture}
     \begin{feynman}
     	\vertex (v1);
     	\vertex[right=1cm] (v2);
     	\vertex[right=2cm of v1] (v3);
     	\vertex[left=.05cm] (v4) {\tiny $x_1$} ;
 		\vertex[right=1cm, below=.05cm of v2] (v5) {\tiny $x_2$};
 		\vertex[right=2.05cm]  (v6) {\tiny $x_3$};
 		\vertex[left=.15cm of v2] (v7);
 		\vertex[right=.15cm of v2] (v8);
     	\diagram*{
     	(v1) -- (v2) -- (v3),
     	(v8) -- [half left, dashed] (v7) -- [half left, dashed] (v8)
     	};
     	\draw[fill=black] (v1) circle(.5mm);
\draw[fill=black] (v2) circle(.5mm);
\draw[fill=black] (v3) circle(.5mm);
     \end{feynman}
 \end{tikzpicture} = \bigg(
 \begin{tikzpicture}
     \begin{feynman}
     	\vertex (v1);
     	\vertex[right=.95cm] (v2);
     	\vertex[left=.02cm] (v3) {\tiny $x_1$} ;
 		\vertex[right=1.0cm] (v4) {\tiny $-$};
     	\diagram*{
     	(v1) -- (v2)
     	};
\draw[fill=black] (v1) circle(.5mm);
     \end{feynman}
 \end{tikzpicture}-\begin{tikzpicture}
     \begin{feynman}
     	\vertex (v1);
     	\vertex[right=.95cm] (v2);
     	\vertex[left=.02cm] (v3) {\tiny $x_1$} ;
 		\vertex[right=1.0cm] (v4) {\tiny $+$};
     	\diagram*{
     	(v1) -- (v2)
     	};
\draw[fill=black] (v1) circle(.5mm);
     \end{feynman}
 \end{tikzpicture}
 \bigg)
 \bigg(
 \begin{tikzpicture}
     \begin{feynman}
     	\vertex (v1);
     	\vertex[right=.95cm] (v2);
     	\vertex[left=.02cm] (v3) {\tiny $-$} ;
 		\vertex[right=1.0cm] (v4) {\tiny $x_2$};
     	\diagram*{
     	(v1) -- (v2)
     	};
\draw[fill=black] (v2) circle(.5mm);
     \end{feynman}
 \end{tikzpicture}-
 \begin{tikzpicture}
     \begin{feynman}
     	\vertex (v1);
     	\vertex[right=.95cm] (v2);
     	\vertex[left=.02cm] (v3) {\tiny $+$} ;
 		\vertex[right=1.0cm] (v4) {\tiny $x_2$};
     	\diagram*{
     	(v1) -- (v2)
     	};
\draw[fill=black] (v2) circle(.5mm);
     \end{feynman}
 \end{tikzpicture} \bigg)
\end{align}
The factorization at the level of the answer is inherited from factorization at the level of the integrand.  Factorization in the answer is manifest as a shuffle product in the symbol and so the contribution is 
\begin{align}
	\frac{X_2+Y_1+Y_2}{X_1+X_2+X_3} \otimes \left[\frac{X_1-Y_1}{X_1+Y_1}\otimes \frac{X_3-Y_2}{X_3+Y_2}+ \frac{X_3-Y_2}{X_3+Y_2}\otimes \frac{X_1-Y_1}{X_1+Y_1} \right]
\end{align}
Now we look at the larger subgraphs, where we have 
\begin{align}
	\begin{tikzpicture}
     \begin{feynman}
     	\vertex (v1);
     	\vertex[right=1cm] (v2);
     	\vertex[right=2cm of v1] (v3);
     	\vertex[left=.05cm] (v4) {\tiny $x_1$} ;
 		\vertex[right=1cm, below=.05cm of v2] (v5) {\tiny $x_2$};
 		\vertex[right=2.05cm]  (v6) {\tiny $x_3$};
 		\vertex[left=.15cm of v1] (v7);
 		\vertex[right=.15cm of v2] (v8);
     	\diagram*{
     	(v1) -- (v2) -- (v3),
     	(v8) -- [out = 120, in = 60, dashed] (v7) -- [out = 300, in = 240, dashed] (v8)
     	};
     	\draw[fill=black] (v1) circle(.5mm);
\draw[fill=black] (v2) circle(.5mm);
\draw[fill=black] (v3) circle(.5mm);
     \end{feynman}
 \end{tikzpicture} = -\bigg(
 \begin{tikzpicture}
     \begin{feynman}
     	\vertex (v1);
     	\vertex[right=.95cm] (v2);
     	\vertex[left=.02cm] (v3) {\tiny $x_1$} ;
 		\vertex[right=1.0cm] (v4) {\tiny $-$};
     	\diagram*{
     	(v1) -- (v2)
     	};
\draw[fill=black] (v1) circle(.5mm);
     \end{feynman}
 \end{tikzpicture}-\begin{tikzpicture}
     \begin{feynman}
     	\vertex (v1);
     	\vertex[right=.95cm] (v2);
     	\vertex[left=.02cm] (v3) {\tiny $x_1$} ;
 		\vertex[right=1.0cm] (v4) {\tiny $+$};
     	\diagram*{
     	(v1) -- (v2)
     	};
\draw[fill=black] (v1) circle(.5mm);
     \end{feynman}
 \end{tikzpicture}+
 \begin{tikzpicture}
     \begin{feynman}
     	\vertex (v1);
     	\vertex[right=.95cm] (v2);
     	\vertex[left=.02cm] (v3) {\tiny $-$} ;
 		\vertex[right=1.0cm, below = .05cm of v2] (v4)  {\tiny $x_2$};
 		\vertex[right=2.0cm] (v5) {\tiny $+$};
     	\diagram*{
     	(v1) -- (v2)--(v5)
     	};
\draw[fill=black] (v2) circle(.5mm);
     \end{feynman}
 \end{tikzpicture}-\begin{tikzpicture}
     \begin{feynman}
     	\vertex (v1);
     	\vertex[right=.95cm] (v2);
     	\vertex[left=.02cm] (v3) {\tiny $+$} ;
 		\vertex[right=1.0cm, below = .05cm of v2] (v4) {\tiny $x_2$};
 		\vertex[right=2.0cm] (v5) {\tiny $+$};
     	\diagram*{
     	(v1) -- (v2) -- (v5)
     	};
\draw[fill=black] (v2) circle(.5mm);
     \end{feynman}
 \end{tikzpicture}
 \bigg) \nonumber\\
 \times \bigg(
 \begin{tikzpicture}
     \begin{feynman}
     	\vertex (v1);
     	\vertex[right=.95cm] (v2);
     	\vertex[left=.02cm] (v3) {\tiny $-$} ;
 		\vertex[right=1.0cm] (v4) {\tiny $x_3$};
     	\diagram*{
     	(v1) -- (v2)
     	};
\draw[fill=black] (v2) circle(.5mm);
     \end{feynman}
 \end{tikzpicture}-
 \begin{tikzpicture}
     \begin{feynman}
     	\vertex (v1);
     	\vertex[right=.95cm] (v2);
     	\vertex[left=.02cm] (v3) {\tiny $+$} ;
 		\vertex[right=1.0cm] (v4) {\tiny $x_3$};
     	\diagram*{
     	(v1) -- (v2)
     	};
\draw[fill=black] (v2) circle(.5mm);
     \end{feynman}
 \end{tikzpicture} \bigg)
\end{align}
The symbol contribution for the two subgraphs is therefore 
\begin{align*}
	\frac{X_1+X_2+Y_2}{X_1+X_2+X_3}\otimes &\bigg[\frac{X_1-Y_1}{X_1+Y_1}\otimes \frac{X_3-Y_2}{X_3+Y_2}+\frac{X_3-Y_2}{X_3+Y_2}\otimes \frac{X_1-Y_1}{X_1+Y_1} \\
	 &\frac{X_2+Y_2-Y_1}{X_2+Y_2+Y_1}\otimes \frac{X_3-Y_2}{X_3+Y_2}+ \frac{X_3-Y_2}{X_3+Y_2}\otimes \frac{X_2+Y_2-Y_1}{X_2+Y_2+Y_1} \bigg]\\\\
	 \frac{X_3+X_2+Y_1}{X_1+X_2+X_3}\otimes &\bigg[\frac{X_3-Y_2}{X_3+Y_2}\otimes \frac{X_1-Y_1}{X_1+Y_1}+\frac{X_1-Y_1}{X_1+Y_1}\otimes \frac{X_3-Y_2}{X_3+Y_2} \\
	 &\frac{X_2+Y_1-Y_2}{X_2+Y_1+Y_2}\otimes \frac{X_1-Y_1}{X_1+Y_1}+ \frac{X_1-Y_1}{X_1+Y_1}\otimes \frac{X_2+Y_1-Y_2}{X_2+Y_1+Y_2} \bigg]
\end{align*}
We combine all these contributions to get the total symbol, which reproduces the 104 term expression in \cite{cosmopoly}.  There does not appear to be a natural set variables which makes the symbol separate into pieces obviously corresponding to the symbols of classical polylogs as with the previous cases.
\section{Integrating the Symbols}\label{integration}
The symbol determines the polylogarithmic function up to functions of lower transcendentality times transcendental numbers.  By integrating the symbol to get the piece of highest transcendental weight, we can then impose boundary conditions to fix the other pieces.  There is no general algorithm for associating a function with a symbol and beyond weight three we lose the guarantee that we only have to consider classical polylogarithms.  Nonetheless, we were able to compute the integrals for the symbols presented in the examples.  We also note that the \textit{Mathematica} package developed in \cite{polylogtools} was of great value in checking or performing various non-trivial symbol computations efficiently. 
\subsection{Two-Site Chain}
In the $u$ and $v$ variables defined earlier the symbol for the two-site chain expands out to
\begin{align}
	\mathcal{S}_2 &= (1-u)\otimes u+(1-v)\otimes v-(1-u)\otimes(1-v)-(1-v)\otimes(1-u)
\end{align}
we immediately recognize these as corresponding to 
\begin{align}
	F_2 = -\text{Li}_2(u)-\li_2(v)-\li_1(u)\li_1(v)+C
\end{align}
It remains to fix the constant.  The correlator should vanish when the edge energy goes to zero.  This subspace is defined by the constraint $u+v = 1$ and so we demand 
\begin{align}
	-\text{Li}_2(u)-\li_2(1-u)-\li_1(u)\li_1(1-u)+C = 0
\end{align} 
from which we deduce $C = \frac{\pi^2}{6}$ due to the Euler identity, which holds for $u, v < 1$, clearly obeyed in the physical region.  Therefore 
\begin{align}
	F_2 = -\text{Li}_2(u)-\li_2(v)-\li_1(u)\li_1(v)+\frac{\pi^2}{6}
\end{align}
which is the result found in \cite{cosmocollider} and again in \cite{cosmobootstrap}.  It is amusing to note that this answer is naturally rearranged using the Abel identity into the following form 
\begin{align}
	F_2 = -\text{Li}_2\left(\frac{X_1-Y}{X_1+Y} \right)-\text{Li}_2\left(\frac{X_2-Y}{X_2+Y} \right)+\text{Li}_2\left(\frac{(X_1-Y)(X_2-Y)}{(X_1+Y)(X_2+Y)} \right)+\frac{\pi^2}{6}
\end{align}
where this holds when $u, v \notin (1, \infty)$.
\subsection{Two-Site Loop}
We recognized that the two-site loop was a sum of two-site chains from which we deduce 
\begin{align}
	F_{\text{1-loop}} = -\text{Li}_2(u_3)-\li_2(v_3)-\li_1(u_3)\li_1(v_3)\nonumber\\
	+\text{Li}_2(u_2)+\li_2(v_2)+\li_1(u_2)\li_1(v_2)\nonumber\\
	+\text{Li}_2(u_1)+\li_2(v_1)+\li_1(u_1)\li_1(v_1)\nonumber\\
	+C
\end{align}
where it remains to fix the constant again.  Setting either of the two edges to zero gives us an equation identical to the two-site chain simply with the opposite sign and so we have 
\begin{align}
F_{\text{1-loop}} = -\text{Li}_2(u_3)-\li_2(v_3)-\li_1(u_3)\li_1(v_3)\nonumber\\
	+\text{Li}_2(u_2)+\li_2(v_2)+\li_1(u_2)\li_1(v_2)\nonumber\\
	+\text{Li}_2(u_1)+\li_2(v_1)+\li_1(u_1)\li_1(v_1)\nonumber\\
	-\frac{\pi^2}{6}
\end{align}
we take up integration over spatial loop momenta in an upcoming work.
\subsection{Three-Site Chain}
\text{}\indent The three-site chain has transcendental weight three.  At weight three it is a theorem that you can integrate a symbol to a function expressed only in terms of classical polylogarithms.  There is an approach to integration which consists of separating the pieces of the symbol based on their symmetry properties under exchange of slots.  This approach was employed succesfully in \cite{goncharov}.  For example, at weight three the only function which can produce a symbol with any antisymmetry  in the last two slots is $\li_2(x)\log(y)$.  We can therefore turn the rightmost tensor symbol into a $\wedge$ and simplify in order to isolate this part of the symbol.  We recognize that it corresponds to two-site chains times logs.   When we have subtracted this piece, we can project onto antisymmetry in the first two slots to separate the $\li_3(z)$ piece, and then finally we are left with $\log^3(w)$'s.  This algorithm still requires some guesswork in practice, but in this case we deduced the answer. We state the answer below  
\begin{align}
	F_{3} = \sum\limits_{\{\sigma_1, \sigma_2\} \in \{-, +\}^2} \sigma_1\sigma_2\bigg[\li_3\left(-\frac{u_{\sigma_1}v_{\sigma_2}}{1-u_{\sigma_1}-v_{\sigma_2}} \right)+\li_3\left(\frac{u_{\sigma_1}}{1-v_{\sigma_2}} \right)+\li_3\left(\frac{v_{\sigma_2}}{1-u_{\sigma_1}} \right)\nonumber\\
	+\li_3\left(\frac{1-u_{\sigma_1}-v_{\sigma_2}}{1-v_{\sigma_2}} \right)+\li_3\left(\frac{1-u_{\sigma_1}-v_{\sigma_2}}{1-u_{\sigma_1}} \right)\nonumber\\
	-\frac{1}{2}\li_1\left(\frac{1-u_{\sigma_1}-v_{\sigma_2}}{1-v_{\sigma_2}} \right)\li_1^2\left(\frac{u_{\sigma_1}}{1-v_{\sigma_2}}\right)-\frac{1}{2}\li_1\left(\frac{1-u_{\sigma_1}-v_{\sigma_2}}{1-u_{\sigma_1}} \right)\li_1^2\left(\frac{v_{\sigma_2}}{1-u_{\sigma_1}} \right) \nonumber\\
	+\frac{1}{6}\li_1^3\left(\frac{v_{\sigma_2}}{1-u_{\sigma_1}} \right)+\frac{1}{6}\li_1^3\left(\frac{u_{\sigma_1}}{1-v_{\sigma_2}} \right)-\frac{1}{6}\li_1^3\left(u_{\sigma_1}+v_{\sigma_2} \right)  \bigg]\nonumber\\
+\li_1\left(1-u_+ \right)\bigg[\li_2\left(\frac{1-v_+-u_+}{1-v_--u_+} \right)-\li_2\left(\frac{v_-(1-v_+-u_+)}{v_+(1-v_--u_+)} \right)\nonumber\\
	-\li_2\left(\frac{1-v_+-u_-}{1-v_--u_-} \right)+\li_2\left(\frac{v_-(1-v_+-u_-)}{v_+(1-v_--u_-)}\right)  \bigg]\nonumber\\
	+\li_1\left(1-v_+ \right)\bigg[\li_2\left(\frac{1-v_+-u_+}{1-v_+-u_-} \right)-\li_2\left(\frac{u_-(1-v_+-u_+)}{u_+(1-v_+-u_-)} \right)\nonumber\\
	-\li_2\left(\frac{1-v_--u_+}{1-v_--u_-} \right)+\li_2\left(\frac{u_-(1-v_--u_+)}{u_+(1-v_--u_-)}\right)  \bigg]\nonumber\\
	+\li_1\left(1-u_+ \right)\li_1\left(1-v_+ \right)\li_1\left(\frac{(v_--v_+)(u_--u_+)}{(1-u_+-v_-)(1-u_--v_+)}\right)
\end{align}
Where we have used the variables 
\begin{align}
\label{gonch}
	u_{\pm} = \frac{X_1\pm Y_1}{X_1+X_2+X_3} \hspace{12mm} v_{\pm} = \frac{X_3\pm Y_2}{X_1+X_2+X_3}
\end{align}
It remains to fix contributions not detected by the symbol.  Setting $Y_1 = 0$ or $Y_2 = 0$ gives $u_+ =  u_-$ or $v_+ = v_-$ respectively.  Demanding that $F_3$ vanish in either of these cases does not require the addition of any new terms.  It is impossible to add a non-trivial logarithm with rational argument that vanishes when either $Y_1$ or $Y_2$ is equal to zero and does not spoil the discontinuity structure of the answer.  This fully fixes the tree-level five-point contribution to be the one stated above.
\subsection{Four-Site and Beyond}
Already at weight four the algorithmic procedure for integrating the symbol by using its symmetry properties is contingent upon restricting to classical polylogarithms, which must be justified.  For a generic integrable symbol, there is no guarantee beyond weight three that it can be produced by a function only consisting of classical polylogarithms.  The authors in \cite{goncharov} were motivated by a theorem first conjectured in \cite{conjecture} to only consider classical polylogarithms.  The statement of the theorem is that a symbol at weight four obeying the following relation under permutation of slots
\begin{align}
	S_{abcd}-S_{bacd}-S_{abdc}+S_{badc}+S_{cdab}-S_{dcab}-S_{cdba}+S_{dcba} = 0
\end{align}
can in fact be integrated to a function consisting only of classical polylogarithms.  We can readily test this conjecture on the wave function at six points using the symbol rule. At six points (weight four) we now have two graph topologies that contribute at tree-level, the four-site chain and the four-site star.  Neither symbol vanishes under this permutation.  It appears that at weight four we lose the capacity to express even the tree-level contributions with classical polylogarithms.  The result of the left-hand side Eq. \ref{gonch} might nonetheless still carry some interesting content and motivate what sorts of polylogarithms to consider at higher multiplicity.  We leave these questions to future work.

\subsection{Comments on General Couplings and Backgrounds}\label{moregeneral}
Though the recursive rule only applies exactly to the each perturbative wave function contribution for the case $\lambda_k(\eta) = \frac{\lambda}{\eta}$ it is still relevant for more general backgrounds.  Consider the case of $\lambda_k(\eta) = \frac{\lambda_k}{\eta^p}$ for any $p \in \Z^+$. In this case, when passing to Fourier space we have 
\begin{align*}
	\lambda_k(\eta) = -i\lambda_k \int\limits_0^{\infty(1-i\epsilon)}(i\varepsilon)^{p-1} e^{i\varepsilon \eta}d\varepsilon
\end{align*}
We have considered the case $p = 1$ thus far.  For $ p > 1$ we get non-trivial numerators multiplying the integrands $\psi_\mathcal{G}$  computed using tools in \cite{cosmopoly}.  For a graph with $N$ vertices, the numerator would be (ignoring powers of $i$)
\begin{align*}
	\mathcal{N} = \prod\limits_{i = 1}^N (x_i-X_i)^{p-1}
\end{align*} 
where the $x_i$ are the energies to be integrated from their corresponding external energies $X_i$ to infinity.  Since the $X_i$ are constant, we already see that there will be a piece computed exactly with the recursive rule but multiplied by the product of external energy sums.  Other pieces will in general produce transcendental functions times polynomials.  A situation of interest is $p = 2$ as it applies to mass insertions in $dS_4$.  We leave a systematic study of general power law backgrounds to future work.

\section{Conclusion and Outlook}
In this paper we have derived a simple recursive rule for computing the symbols of perturbative contributions to the vacuum wave function.  The rule applies exactly to a case of particular interest, $\lambda \phi^3$ in $dS_4$.  This case is interesting for reasons both concrete and speculative.  It was demonstrated \cite{cosmobootstrap} that the four-point function of conformally coupled scalars exchanging a general massive scalar is sufficient to systematically generate four-point functions with massless external legs exchanging fields of general spin.  Our case considered exchange of a conformally coupled scalar, from which one can obtain correlators of massless external legs exchanging conformally coupled fields with spin, which includes gauge fields and partially massless fields of spin 2.   It would be interesting to explore generalizing the cosmological bootstrap tools to the five-point function computed here. The results presented here also apply to the computation of Witten diagrams in Euclidean AdS.  It would be interesting to explore whether symbol recursion has a nice boundary interpretation in this setting, as the rule is manifest in momentum-space, an as yet unexplored setting for 3D Euclidean CFT.   This specific case is also interesting due to its connection to flat-space physics.  A differential operator acting on the total energy discontinuity of a given wave function contribution produces the flat space scattering amplitude.  It is known that a very rich geometric and algebraic structure describes $\lambda \phi^3$ amplitudes in flat space and is connected to the geometry of the open string moduli space.  It would be surprising if this structure only existed on the codimension one subspace of vanishing total energy.  The recursive rule also leaves something to be desired: integration beyond the symbol.  Even at five-points, integration proved to be a much more formidable task than its trivial counterpart at four-points.  Nonetheless, some recursive structure is still clearly present in the answer, with factorization into three and four-point functions manifest.  It would be interesting to explore whether a different basis of polylogs and choices of variables made a generalization to higher multiplicity manifest. 
 \section{Acknowledgments}
 We thank Nima Arkani-Hamed for advising and many helpful discussions throughout this work.  We would also like to thank Akshay Yelleshpur for helpful discussions and comments on the draft.

\end{document}